\documentclass[a4,conference]{IEEEtran/IEEEtran}
\def\ps@headings{
\def\@oddhead{\mbox{}\scriptsize\rightmark \hfil \thepage}
\def\@evenhead{\scriptsize\thepage \hfil \leftmark\mbox{}}
\def\@oddfoot{}
\def\@evenfoot{}}
\makeatother

\usepackage{pgfplots}
\usepackage{color}
\usepackage{subfig}
\usepackage{graphicx}
\usepackage{cite}
\usepackage{amssymb,amsfonts}
\usepackage{url}
\usepackage[utf8]{inputenc}
\usepackage{tabularx}
\usepackage{hhline}
\usepackage{multirow}
\usepackage{wrapfig}
\usepackage{pifont}
\usepackage{comment}

\usepackage[outline]{contour}

\usepackage{pifont}
\usepackage{tikz}
  \usetikzlibrary{arrows,shapes,calc,positioning,shadows,trees,patterns,decorations}
  \tikzset{
    ncbar angle/.initial=90,
    ncbar/.style={
        to path=(\tikztostart)
        -- ($(\tikztostart)!#1!\pgfkeysvalueof{/tikz/ncbar angle}:(\tikztotarget)$)
        -- ($(\tikztotarget)!($(\tikztostart)!#1!\pgfkeysvalueof{/tikz/ncbar angle}:(\tikztotarget)$)!\pgfkeysvalueof{/tikz/ncbar angle}:(\tikztostart)$)
        -- (\tikztotarget)
    },
    ncbar/.default=0.5cm,
}
\tikzset{round left paren/.style={ncbar=0.5cm,out=110,in=-110}}
\tikzset{round right paren/.style={ncbar=0.5cm,out=70,in=-70}}

\usepackage{hyperref}
\usepackage{url}

\usepackage{amsmath}
\usepackage{amsmath,amsthm}
\usepackage{lmodern}
\usepackage{pgfplots}
\usepackage{algorithm}
\usepackage[noend]{algpseudocode}
\usepackage[export]{adjustbox}
\usepackage{soul}

\begin{document}

\title{BlockHouse: Blockchain-based Distributed Storehouse System}

\author{\IEEEauthorblockN{Doriane Perard\IEEEauthorrefmark{1},
Lucas Gicquel\IEEEauthorrefmark{2} and J\'er\^ome Lacan\IEEEauthorrefmark{1}}

\IEEEauthorblockA{\IEEEauthorrefmark{1} ISAE-Supaero, Universit\'{e} de Toulouse, France \\
\IEEEauthorrefmark{2} Edokial, Causse Comtal, 12340 Bozouls \\
Email: \IEEEauthorrefmark{1}firstname.name@isae-supaero.fr \\
\IEEEauthorrefmark{2}gicquel.lucas@edokial.com}

}

\maketitle

\begin{abstract}
%
We propose in this paper \textit{BlockHouse}, a decentralized/P2P storage system fully based on private blockchains. Each participant can rent his unused storage in order to host data of other members. 
This system uses a dual Smart Contract and Proof of Retrievability system to automatically check at a fixed frequency if the file is still hosted. 
In addition to transparency, 
the blockchain allows a better integration with all payments associated to this type of system ( regular payments, sequestration to ensure good behaviors of users, ...). Except the data transferred between the client and the server, all the actions go through a smart contract in the blockchain in order to log, pay and secure the entire storage process.



\end{abstract}

\section{Introduction}
\label{sec:intro}

To face the huge increase of data generated by applications, enterprises must manage their storage by taking into account the reliability, confidentiality and privacy.

Local storage is the first possibility and is optimal for confidentiality and privacy. However, despite the numerous existing local storage systems, it is not 
really easy to 
ensure the full system reliability. 
As a consequence, small or medium enterprises often choose to focus on their main activity and externalize this kind of service.

In this case, the most evident solution is using cloud storage which is cheap and reliable. 
This main issue of this choice is probably the confidentiality and privacy. Indeed, even if the data can be locally encrypted before being stored on the remote cloud, this solution is generally not recommended for critical data. 

A potential solution would be to collaborate with trusted associate enterprises to share the storage resources in a kind of peer-to-peer network. In other words, companies having unused storage capacity can rent this to other enterprises. Technically, this type of service can be supported by an extensive literature on both reliability and confidentiality. 

One of the potential issues in this distributed system is that all the system members must agree on the amount, duration and storage reliability of each file. One of the system members could ensure the role of trust third party by logging all the file exchanges and transactions in a ledger, but this could be problematic for companies storing critical data.

The recent developments of blockchains have shown that such systems can provide trust in a competitive system without trusted third party. The idea of integrating this tool in a distributed storage system is quite natural. However, companies want to control the access to their data and so, traditional public blockchains like Bitcoin or Ethereum, where anonymity is omnipresent, can not be used. These are the reasons why private blockchains have been created: indeed, these blockchains manage the users' account and check the user's identity at each connection. This kind of blockchain is also more efficient than the public ones: they require only few seconds and few computation power to create a block. 

Our proposal is a distributed storage system managed on a private blockchain. 
A storage smart contract contains the amount of data and the storage duration.
In addition to transparency, this allows a better integration with all payments associated to this type of system (regular payments, sequestration to ensure good behaviors of the users, ...). 
The system also periodically checks that data is really stored on the servers by using cryptographic tools, called Proof of Retrievability (PoR) \cite{tan2018survey} that will be stored on the blockchain as well.
Except the data transferred between the client and the server, all the actions go through a smart contract in the blockchain in order to log, pay and secure the entire storage process.

The paper is organized as follows. We present the related works in section \ref{sec_rel_work}. All the details about \textit{BlockHouse} are presented in section \ref{sec_blockhouse_description}. We propose some extensions in section \ref{sec:Discussion} before concluding in section \ref{sec:conclusion}.

\section{Related works}
\label{sec_rel_work}

In this Section, we first introduce the key concepts of PoR which is one of the main components of our system. Then we will present other works using blockchains and storage systems.

\subsubsection{Proof of retrievability}
\label{sec:proofOfStorage}
is a mathematical tool that allows a server to prove to a client that it does have specific data, that can be fully recovered. The client sends random challenges, and the server answers with a concurring proof. A proof is much smaller than the original file, therefore it can be frequently transferred. To verify it, the client needs the metadata, a small amount of data computed from the original file. It is thus possible to be sure that the file is still stored on the remote server, without having to download entirely.


These concepts were introduced by Juels and Kaliski \cite{juels2007pors}.
The survey \cite{tan2018survey} is a good overview of existing proofs of retrievability classified according to different attributes.

\subsubsection{Distributed storage and blockchains}
\label{sec:otherStorageBlockchains}
Since blockchains are managed by a peer-to-peer network, it is quite natural to integrate it in a distributed storage system. Storj, BlockStore and Sia are 3 different systems that propose P2P storage using blockchain.

Sia \cite{sia} is a Bitcoin-based blockchain aiming at storing client files on several hosts. The contracts between the clients and the hosts are stored on the blockchain. The file is first split into encrypted chunks and encoded with erasure codes~\cite{erasureCodeSurvey:2013}. The hosts are asked to provide regular proofs-of-storage (with a limited number of possible challenges). However, in such a public system, the end of the contract just relies on the reputation and on the file redundancy which incites the host to return the file to the client.

The Storj system \cite{wilkinson2014storj} uses similar concepts and suffers from the same drawback to not provide guarantees on the file recovery at the end of the contract. This is not acceptable for companies.



BlockStore \cite{ruj2018blockstore} is one of the first to detail the overall scheme of the file storage process. It sets up all the advanced storage system basics. It uses \textit{Space Wallet}, a special structure that tracks available storage space on all nodes. Their technical choice is to reduce the blockchain weight by storing only failing proofs in the blockchain. The other features (audits, transfer, ...) are off-chain. However, it does not solve the file recovery issue at the end of the storage.

\section{BlockHouse description}
\label{sec_blockhouse_description}

This section presents our proposal \textit{BlockHouse}, which can be decomposed into three steps: The \textit{initialization} which mainly includes the negotiation of the storage contract, then the \textit{storage and audit} part which contains the periodic proofs-of-retrievability and the corresponding payments, and the \textit{end of the contract} which is the most critical with the final payment and the client file restitution. 

\subsection{Initialization step}

\begin{figure}
  \includegraphics[width=\linewidth]{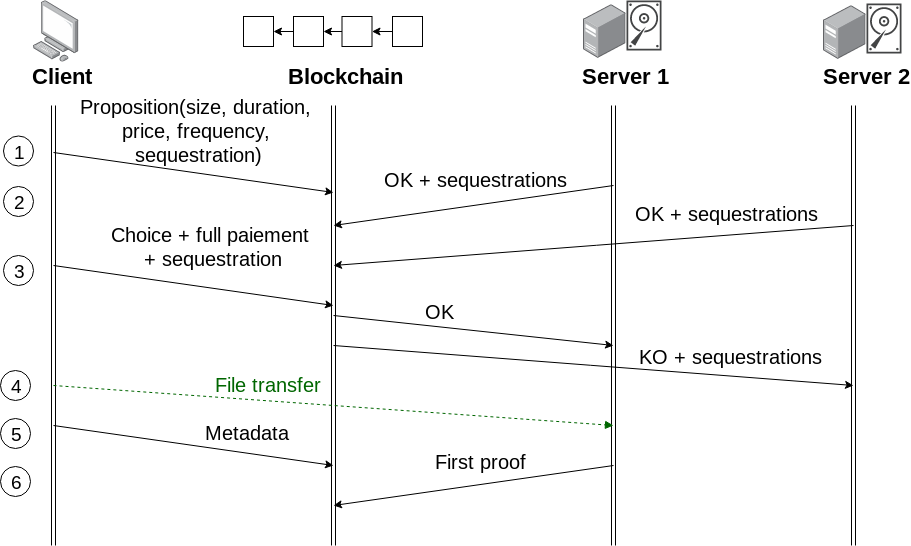}
  \caption{Initialization step}
  \label{fig:step1}
\end{figure}


Fig. \ref{fig:step1} presents the initialization step of the \textit{BlockHouse} protocol.

\subsubsection{Contract creation}
A node (the client) which wants to rent some storage can make a proposition to the network. To do this, he makes a special transaction (\ding{192} in the Fig.) with some details about his data (the data size, the contract duration, the proof of retrievability frequency wanted, the price, the file sequestration needed). Every participant in the blockchain network is able to know who wants what.

Thanks to a reputation score, based on the blockchain history, every node can be rated on different criteria: on one hand, the clients are rated on their number of litigation, on the other hand, hosts are rated on the ratio of succeed proofs to submitted proofs.  When a potential host with enough free storage wants to host the data, it makes a transaction \ding{193} to request an agreement with a client he found sufficiently rated. When a client has chosen the best host, it makes a last transaction \ding{194} to validate the agreement. It can choose several hosts to improve availability. It finally transfers the full payment of the file storage to the smart contract.


The client sends the data to the host(s) (\ding{195}) and generates the metadata for the future proofs of retrievability (\ding{196}).

This initialization phase ends when the host sends its first proof of retrievability \ding{197}.

\subsubsection{Sequestration}

To assure that both participants honestly play the game, they must deposit some tokens in a smart contract in \ding{193} and \ding{194}. These tokens are a kind of sequestration, to push them to respect the contract until the end. There are two types of sequestration, with different goals. Both are deposited by the storage server when he responds positively to a proposal.
\begin{itemize}
    \item The first sequestration, called \textit{file sequestration}, is sent to the client if the hosting server is not able to return the original file.
    \item The second sequestration is used in case of conflict at the end of the contract. Indeed, if the client declares that he can not download the file back, it is not possible to be sure that the client is honest or not. In this case, we have to ask to other nodes to check, and they are rewarded with the dishonest node's sequestration. These tokens will be called \textit{auditors sequestration}, and both parties have to deposit it.
\end{itemize}

\subsection{Storage and audit part}
\begin{figure}
  \includegraphics[width=\linewidth]{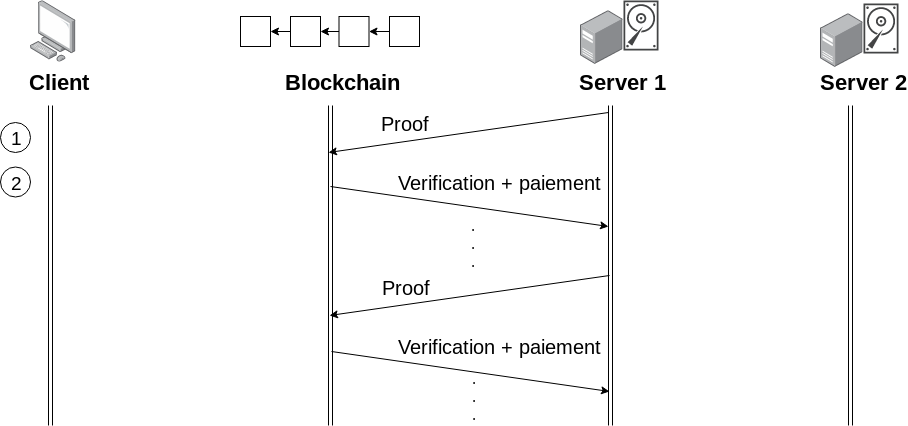}
  \caption{Audits}
  \label{fig:step2}
\end{figure}

Fig. \ref{fig:step2} presents the second part of the \textit{BlockHouse} protocol, which is repeated according to the announced frequency.

Once these formalities have been completed, the host has to frequently perform proof of retrievability  to be payed. The way we propose to implement the time in the blockchain is discussed in \ref{sec:trigger}. 


The challenge is randomly generated from the blockchain itself, as explained in \ref{sec:randomData}. Thus, it is deduced directly from the system and not required to be chosen by one node and sent to the others, saving storage space and bandwidth.

Every time the host performs a proof of retrievability, it includes it in a new transaction (\ding{193}). A smart contract receives this transaction, and check the proof using the initial data. If this proof is asked (the frequency is good) and correct, a payment is sent from the smart contract to the host (\ding{193}). 

\subsection{End of the contract}
\label{sec:EndOfTheContract}
\begin{figure}
  \includegraphics[width=\linewidth]{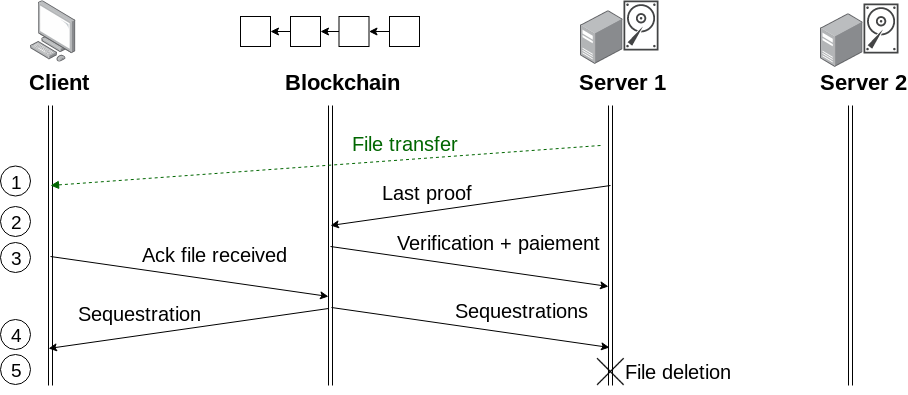}
  \caption{End of the contract}
  \label{fig:step3}
\end{figure}

Fig. \ref{fig:step3} presents the last part of the \textit{BlockHouse} protocol, with a classic end (without issue).

In common use, just before the end, the client has to download its file \ding{192}. In the meantime, the host continue to create proofs of retrievability \ding{193}. If the downloaded file is correct, the client announces it in a special transaction \ding{194}. Then, the sequestrations are given back to their owners (\ding{195}) and the file is deleted (\ding{196}).

However, there are other reasons why the contract could be interrupted:
\begin{itemize}
    \item Early termination by the client: in the event that the client decides to early terminate the contract, each party recovers its sequestration, and the money used to pay the proofs is fully sent to the server.
    \item Early termination by the hosting server: in the event that the server decides to early terminate the contract, the client recovers his sequestration and money used to reward the proof, but also the file sequestration from the server. The server recovers its auditors sequestration. 
    \item Incorrect proof of storage: if there is too much incorrect proofs of storage from the hosting server, the contract is terminated. As the previous case, the client recovers his sequestration and money used to reward the proofs, but also the file sequestration from the server. The server recovers its auditors sequestration. 
    \item Ending download issue: this is the most complex case. The customer must issue a correct termination message at the end of the contract, when it has successfully downloaded its file. In the event that it sends an incorrect message, it is not possible for the rest of the network to judge which one of the two parties is dishonest.
    Indeed, the client may not succeed in downloading the file because the server does not make it available, or may try to cheat to receive the file sequestration from the server. 
    To decide between the two, we use others nodes, called auditors. 
\end{itemize}

\subsection{Auditors}
Since the file is downloaded outside the blockchain, in case of disagreement it is impossible for a network user to know who is right. To know if the client is lying (and therefore the file is available) or if the server is lying (the file is not available), other nodes, called \textit{auditors}, are asked to check. The chosen auditors who correctly answer (i.e. in agreement with the majority) are rewarded with a part of the auditors sequestration.

Since verification is time-consuming, it may be useful to implement sanctions, for example banning a user who is at fault, in order to encourage everyone to act honestly. This is possible because we are in the case of a private blockchain, where nodes are identified and accepted in the network.


Auditors must be selected randomly from the blockchain and in sufficient quantity to ensure a majority of honest response, as we explain in \ref{sec:proba}.

\section{Discussion}
\label{sec:Discussion}
\subsection{Random data (seed)}
 \label{sec:randomData}
As explained previously, we want the blockchain to randomly generate random challenges and choose auditors.

In the blockchain, the too obvious way would be to use the block hash as a seed for a pseudo-random number generator. However, a party can control the hash of a block to predict the future block hash \cite{chatterjee2019probabilistic} and then control the future challenges. 


Other ways to generate random numbers would be using multiple blockchain elements like the $i$-th block hash instead of using the last block hash, smart contract address, timestamp, client or server address, last proof hash, etc.




\subsection{Regular trigger}
\label{sec:trigger}

To make sure that every host still hosts the clients' files, the system itself will incite them to create a proof at a specific and fixed frequency. Working with a private blockchain, it is not possible to use the block number as fixed frequency: indeed, in this context, blocks are only created when new transactions are processed. We can only work with a clock.

As we can see in the block validation algorithm section of \cite{blockprotocol}, blocks are validated if each nodes' clock in the network are synchronized with at most 15 minutes difference. Thanks to this feature, it is possible to use the nodes' clock in the smart contract to determine the date of every proof.

\subsection{Probability to have a majority of honest auditors}

\label{sec:proba}
\begin{figure}
\begin{tikzpicture}
\centering
    \begin{semilogyaxis}[
        font=\sffamily,
        height=5cm,
        width=\linewidth-0.5cm,
        axis x line=bottom,
        axis y line=left,
        ymajorgrids,
        yminorgrids = true,
        ytick={0,0.000000001,0.00000001,0.0000001,0.000001,0.00001,0.0001,0.001,0.01,0.1,1},
        xlabel = {Number of auditor nodes $n$} ,
		ylabel = {Proba of dishonest majority},
		xmin = 30, xmax = 300, 
		ymin = 0.000000001, ymax = 1 ]
        \addplot+[mark=none, line width=2.25pt] table[x index=0, y index=1] {data/loi_normale_0,6.data};
        \addplot+[mark=none, line width=1.25pt] table[x index=0, y index=1] {data/loi_normale_0,71.data};
        \addplot+[teal, mark=none, line width=0.25pt] table[x index=0, y index=1] {data/loi_normale_0,8.data};

    \end{semilogyaxis}
    \begin{scope}[shift={(0,2)}]
    	\draw[yshift=2cm,xshift=4.75cm, blue] (0,0) 
			plot[mark=., mark options={fill=blue}] (0.25,0) 
			node[right]{\textbf{p = 2/3}};
		\draw[yshift=2cm,xshift=2.25cm,red] (0,0) 
			plot[mark=.] (0.25,0)
			node[right]{p = 5/7};
		\draw[yshift=2cm,xshift=0cm, teal] (0,0) 
			plot[mark=., mark options={fill=teal}] (0.25,0) 
			node[right]{\emph{p = 4/5}};
	\end{scope}
\end{tikzpicture}
	\caption{Probability to have a dishonest majority, depending on $n$, with $p=4/5$, $p=5/7$ and $p=2/3$.}
	\label{fig:proba}
\end{figure}
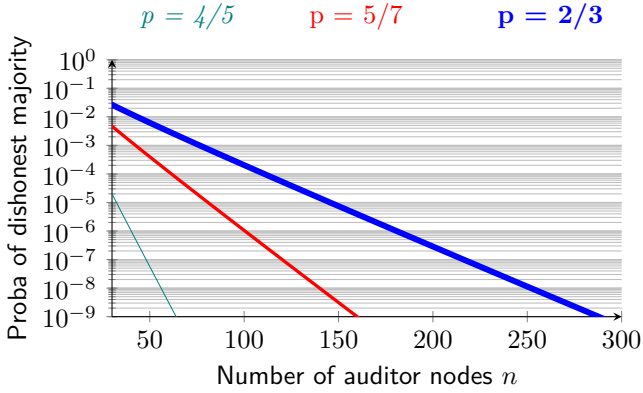

To evaluate the probability to have a honest majority, let us denote $n$ the number of nodes chosen to be auditors. We note $p$ the probability one node responds correctly, and $X_n$ the total number of correct answers. 

We assume that the $X_n$ can be modeled as realizations of an independent discrete random variable following a probability binomial distribution with probability $p$. 

We need a minimum number of nodes in the network for the blockchain to work properly, and for the consensus to be valid. 
We therefore assume that this number is sufficient to be able to approximate the binomial distribution by a normal distribution ($n \geq 30$, $np \geq 5$  and $n(1-p) \geq 5$):

\begin{equation}
    f(x) = \frac{e^{-\frac{1}{2}(\frac{x-\mu}{\sigma})^2}}{\sigma\sqrt{2\pi}} = \frac{e^{-\frac{1}{2}(\frac{x-np}{\sqrt{np(1-p)}})^2}}{\sqrt{np(1-p)}\sqrt{2\pi}}
\end{equation}
The probability to have a dishonest majority is:
{\small \begin{eqnarray} 
\overline{P_m} = p \left( 0 \leq X_n \leq \frac{n}{2} \right) = \int_{0}^{\frac{n}{2}} f(x) dx = \int_{0}^{\frac{n}{2}} \frac{e^{-\frac{1}{2}(\frac{x-np}{\sqrt{np(1-p)}})^2}}{\sqrt{np(1-p)}\sqrt{2\pi}}
\label{eq:normal_law} 
\end{eqnarray}}

In Fig. \ref{fig:proba}, we represent the probability to have a majority of dishonest auditors depending on the number auditor nodes.

The probability is lower than $10^{-6}$ when there are at least 41 nodes with $p=4/5$, 101 nodes with $p=5/7$ and 181 nodes with $p=2/3$.

\section{Conclusion}
\label{sec:conclusion}
The increasing popularity of blockchains reveals all the possibilities a decentralized ledger can offer. Nonetheless blockchains can also have other usages.

In this paper, we introduce a new usage of blockchains in a private context, with the aim of proposing a new decentralized storage system. The \textit{BlockHouse} system is based on a protocol including three main steps done on-chain: the storage contract initialization, regular audits and the end of the contract. It is mainly based on cryptographic proofs of retrievability which allow smart contracts to verify that hosts are storing the data correctly. This system allows to extend the trust on the data stored by the servers thanks to the blockchain.

The major problem that could arise in our system would be that the blockchain size increases significantly and then becomes complicated to store. In order to fix that problem, erasure codes can be used as a way to cancel this growth~\cite{perard2018erasure}.

\section*{Acknowledgement}
We thank Yann Bachy for his comments that greatly improved the paper.


\bibliographystyle{IEEEtran}
\bibliography{main}

\end{document}